\documentclass[twocolumn,prc,aps,showpacs,amsmath,amssymb,nofootinbib]{revtex4}
\usepackage{graphicx}

 \def\Pom{{ I\!\!P}}
 \def\Reg{{ I\!\!R}}
 \def\gsim{\mathrel{\rlap{\lower4pt\hbox{\hskip1pt$\sim$}}
 \raise1pt\hbox{$>$}}}

 \newcommand\la{\langle}
 \newcommand\ra{\rangle}
 \newcommand\beq{\begin{equation}}
 
 \newcommand\eeq{\end{equation}}
 \newcommand\beqn{\begin{eqnarray}}
 \newcommand\eeqn{\end{eqnarray}}
\def\mb{\,\mbox{mb}}
\def\fm{\,\mbox{fm}}
\def\GeV{\,\mbox{GeV}}

\def\lsim{\mathrel{\rlap{\lower4pt\hbox{\hskip1pt$\sim$}}
    \raise1pt\hbox{$<$}}}         
\def\gsim{\mathrel{\rlap{\lower4pt\hbox{\hskip1pt$\sim$}}
    \raise1pt\hbox{$>$}}}         
\def\Re{\,{\rm Re}\,}
\def\Im{\,{\rm Im}\,}
\def\mb{\,\mbox{mb}}
\def\fm{\,\mbox{fm}}
\def\GeV{\,\mbox{GeV}}
\def\MeV{\,\mbox{MeV}}

\begin{document}
\date{}

\title{\bf Pion structure function at small x from DIS data}

\author{B.Z.~Kopeliovich$^{a}$}
\author{I.K.~Potashnikova$^{a}$}
\author{B.~Povh$^b$}
\author{Ivan~Schmidt$^a$}

\affiliation{$^a$Departamento de F\'{\i}sica, Centro de Estudios
Subat\'omicos, Universidad T\'ecnica Federico Santa Mar\'{\i}a,
and\\
Centro Cient\'ifico-Tecnol\'ogico de Valpara\'iso,\\
Casilla 110-V, Valpara\'iso, Chile\\
$^b$Max-Planck-Institut f\"ur Kernphysik, Postfach 
103980, 69029 Heidelberg, Germany
}

\date{\today}

\begin{abstract}
Production of leading neutrons in DIS is usually considered as a tool 
to measure the pion structure function at small $x$ accessible at HERA. 
The main obstacle is the lack of reliable evaluations of the absorption corrections, 
which significantly suppress the cross section. 
We performed a parameter free calculation within the dipole approach and found the 
absorption corrections to be nearly as strong, as for neutron production in $pp$ collisions.
We also included the significant contribution of the iso-vector Reggeons with natural ($\rho$, $a_2$) and unnatural ($a_1$, $\rho$-$\pi$ cut) parity with parameters constrained by phenomenology.
With a certain modeling for the pion-to-proton ratio of the structure functions we reached good agreement with data from the ZEUS and H1 experiments, successfully reproducing the observed dependences on the fractional neutron momentum $z$, the photon virtuality $Q^2$, and the transverse momentum transfer $q_T$.

\end{abstract}


\maketitle

\section{Introduction}\label{intro}

Neutron production in deep-inelastic scattering (DIS) on a proton
can serve as a sensitive tool to study the properties of the meson cloud of nucleons,
because only iso-vector quantum numbers in the crossed channel are allowed. 
If neutrons are produced at forward rapidities with small transverse momenta,
the contribution of large impact parameters of $\gamma^*p$ collisions dominates,
so one can probe light mesons in the proton wave function, in particular pions.
In terms of the dispersion relation this means that in this kinematic region one gets 
close to the pion pole.

\begin{figure}[htb]
\centerline{
  \scalebox{0.35}{\includegraphics{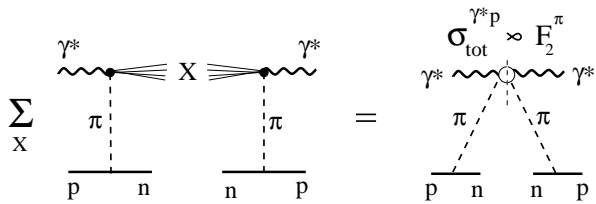}}}
\caption{\label{pion} Graphical representation of the cross
section of inclusive neutron production in hadron-proton collisions,
in the fragmentation region of the proton. }
 \end{figure}

Thus, one can treat leading neutron production in DIS as a method to measure 
the structure function of the pion,
$F_2^\pi(x_\pi,Q^2)$, as is illustrated in Fig.~\ref{pion}.
 Summing up all final states $X$ at a fixed invariant mass $M_X$
  one arrives at the total hadron-pion
cross section at c.m. energy $M_X$. This cross section is a slowly
varying function of $M_X$, what leads
to an approximate Feynman scaling. 
The rapidity gap covered by the pion exchange correspond to the energy, which
is much smaller than the total c.m. collision energy squared

 \beq
{s'\over s_0}={s\over M_X^2}\approx {1\over 1-z}\,,
\label{50}
 \eeq
 where $s_0$ is the scale factor, usually fixed at $1\GeV^2$; and
$z=p_n^+/p_p^+$ is the fraction of the proton light-cone momentum
carried by the neutron. If $z$ is large, it is close to Feynman $x_F$.

The pion exchange brings in the cross section the factor $(1-z)^{-2\alpha_\pi(t)}$, where
$\alpha_\pi(t)$ is the pion Regge trajectory.
This factor is independent of the collision energy, if $z$ is fixed,
so the pion exchange contribution does not vanish with energy. The smaller is the
4-momentum transfer squared $t$, the closer one approaches the pion
pole in the dispersion relation, and the more important is the pion contribution. 
However, the smallest
values of $t$ are reached in the forward direction at $z\to1$.
The latter condition leads to the dominance of other
Reggeons which have higher intercepts. Indeed, the corresponding
Regge factor $(1-z)^{-2\alpha_{\footnotesize\Reg}(t)}$ for $\rho$ and
$a_2$ Reggeons is about $1/(1-z)$ times larger than the one for
pion. Although in general these Reggeons are suppressed by an order
of magnitude compared to the pion \cite{k2p}, they become
important at $z\gsim0.9$.

The effective contribution of the axial-vector states ($a_1$ pole and $\rho$-$\pi$ cut) with the parameters fixed from phenomenology, 
was found recently \cite{kpss-spin} to be 
crucial explaining data on azimuthal asymmetry of leading neutrons produced in $pp$ collisions.
This Reggeon having a low intercept, affects the cross section at small $z$.

The most important correction, which is the main focus of this paper,
is the effect of absorption, or initial/final state interactions.
The active projectile partons participating in the reaction, as well
as the spectator ones, can interact inelastically with the proton target or with
the recoil neutron, and initiate particle production, which usually
leads to a substantial reduction of the fractional neutron momentum. The
probability that this does not happen, called sometimes survival
probability of a large rapidity gap, leads to a suppression of
leading neutrons produced at large $z$, because this process is associated with
formation of a rapidity gap $\Delta y\sim-\ln(1-z)$. Some calculations
predict quite a mild effect, of about $10\%$ even in the soft process $pp\to nX$ 
\cite{boreskov1,boreskov2,3n,ap}, while others
\cite{strong2,kkmr,kmr,kpss} expect a strong reduction by about a
factor of 2. See \cite{kkmr} for a discussion of the current
controversies in data and theory, for leading neutron production.

Notice that usually the absorptive corrections are calculated in a probabilistic
way, convolving the gap survival probability with the cross
section. We found, however, that the spin amplitudes of neutron
production acquire quite different suppression factors \cite{kpss,kpss-spin}, and one
should work with amplitudes, rather than with probabilities.

At first glance the absorptive corrections of the hadronic fluctuations of a highly virtual photon
should be vanishingly small. However, the observed weak $Q^2$ dependence of nuclear shadowing in DIS
demonstrates that this is not true: both shadowing and absorption are dominated by rare soft fluctuations of the photon \cite{interplay}. This is why the absorptive corrections were calculated in \cite{ap} relying on the effective
absorption cross section adjusted to data on nuclear shadowing \cite{universal}.

Even more simplified evaluation of absorptive corrections were performed in \cite{kkmr,kmr},
basing on the two component model for fluctuation of the virtual photon, soft and hard. The former was assumed to interact like the $\rho$ meson, while the latter cross section was fixed zero.

Below we perform explicit calculations of the absorptive corrections caused by 
the interactions of the fluctuations of a highly virtual photon  within the dipole approach cite{zkl}.
Moreover, like in $pp$ collisions \cite{kpss}, even a stronger absorption, related to the formation of a large color octet dipole  in $\gamma^*\pi$ interaction, affects the large-$z$ part of the neutron spectrum.
This effect has been missed in previous calculations of the absorption corrections.

Below our results are presented as follows. In Sect.~\ref{born} the spin structure of the amplitude without absorption corrections is presented, and the theoretical uncertainties in the evaluation of the cross section are discussed. 
Although the goal of the present paper is to study the possibilities of extraction of the pion structure function from data, we try to predict the cross section of leading neutron production, modeling the ratio of pion to proton structure functions. 

Sect.~\ref{absorption} is devoted to the absorptive corrections, which are the main focus of this paper.
The important observation is the production of a large size color octet dipole formed by the remnants of the virtual photon and pion. Initial/final state interaction of such a dipole, controlling the absorptive corrections at large $z$,
only slightly depend on on the size of the $\bar qq$ fluctuation of the virtual photon, therefore almost no $Q^2$ is predicted. An examples of the absorption suppression factor $S_{4q}(b)$ and corrected for gluon radiation $\tilde S_{4q}(b)$ are shown in Fig.~\ref{fig:S(b)}. The same figure demonstrates the reduced suppression factor $S_{\gamma^*}(b)$ at smaller $z$, where absorption of the hadronic fluctuations of the virtual photon plays major role.
A sizable $Q^2$ dependence is predicted at smaller $z$, which is confirmed by data. The cross section of leading neutron production is found to be about twice smaller than the absorption uncorrected one, as is demonstrated in Fig.~\ref{fig:abs-corr}.

The isovector Reggeons, which also contribute to the neutron production, are evaluated in Sect.~\ref{reggeons}.
The high-intercept $\rho$-Reggeon is important at large $z$ and large momentum transfer $q_T$ (it flips helicity).
The low intercept $a_1$-Reggeon contributes at smaller $z$. The Regge $a_1$-pole itself is found to be very weak, and is replaced by an effective pole $\tilde a_1$, which also represents the $\rho$-$\pi$ Regge cut.

All the contributions together, corrected for absorption, reproduce data pretty well, as is demonstrated for the $q_T$- and $Q^2$-integrated cross sections in Fig.~\ref{fig:qt-integrated}. The $Q^2$ and $q_T$ dependences are also well reproduced, as is shown in Fig.~\ref{fig:Q2-dep} and \ref{fig:qt-dep} respectively.

The possibility of extraction of the pion structure function from data is discussed in Sect.~\ref{F2pion}, and the sensitivity of the final results to the value of the pion-to-proton ratio of $F_2$ is demonstrated in Fig.~\ref{fig:Np}.

\section{Born approximation}\label{born}

\subsection{Measuring the pion structure function}\label{pion.pole}

In the Born approximation the pion exchange contribution to the amplitude
of neutron production $\gamma^*p\to Xn$, depicted in Fig.~\ref{pion}, in
the leading order in small parameter $m_N/\sqrt{s}$ has the form
 \beq
A^B_{p\to n}(\vec q,z)=
\bar\xi_n\left[\sigma_3\, q_L+
\frac{1}{\sqrt{z}}\,
\vec\sigma\cdot\vec q_T\right]\xi_p\,
\phi^B(q_T,z)\,,
\label{100}
 \eeq
 where $\vec\sigma$ are Pauli matrices;  $\xi_{p,n}$ are the proton or
neutron spinors;  $\vec q_T$ is the transverse component of the momentum transfer;
 \beq
q_L=\frac{1-z}{\sqrt{z}}\,m_N.
\label{110}
 \eeq

 At small $1-z\ll1$ the pseudoscalar amplitude $\phi^B(q_T,z)$ has
the
triple-Regge form,
 \beqn
\phi^B(q_T,z)&=&\frac{\alpha_\pi^\prime}{8}\,
G_{\pi^+pn}(t)\,\eta_\pi(t)\,
(1-z)^{-\alpha_\pi(t)}
\nonumber\\ &\times&
A_{\gamma^*\pi^\to X}(M_X^2)\,,
\label{120}
 \eeqn
where $M_X^2=(1-z)s$, and  the 4-momentum
transfer squared $t$ has the form,
 \beq
-t=q_L^2+{1\over z}\,q_T^2\,,
\label{130}
 \eeq
 and $\eta_\pi(t)$ is the phase (signature) factor which can be expanded near
the pion pole as,
 \beq
\eta_\pi(t)=i-ctg\left[\frac{\pi\alpha_\pi(t)}{2}\right]\approx
i+\frac{2}{\pi\alpha_\pi^\prime}\,
\frac{1}{m_\pi^2-t}\,.
\label{140}
 \eeq
 We assume a linear pion Regge trajectory
$\alpha_\pi(t)=\alpha_\pi^\prime(t-m_\pi^2)$
with $\alpha_\pi^\prime\approx 0.9\GeV^{-2}$.
The imaginary part in (\ref{140}) is neglected in what follows, because its 
contribution near the pion pole is small.

The effective vertex function
$G_{\pi^+pn}(t)=g_{\pi^+pn}\exp(R_1^2t)$, where  
$g^2_{\pi^+pn}(t)/8\pi=13.85$. The value of the slope parameter $R_1$ is specified below.

The amplitudes in (\ref{100})-(\ref{120}) are normalized as,
 \beq
\sigma^{\gamma^*\pi^+}_{tot}(s'=M_X^2)={1\over M_X^2}
\sum\limits_X|A_{\gamma^*\pi^+\to X}(M_X^2)|^2\,,
\label{144}
 \eeq
 where different hadronic final states $X$ are summed at fixed invariant
mass $M_X$. Correspondingly, the differential cross section of inclusive
neutron production reads \cite{bishari,2klp},
 \beqn
z\,\frac{d\sigma^B_{p\to n}}{dz\,dq_T^2}&=&
\left(\frac{\alpha_\pi^\prime}{8}\right)^2
|t|G_{\pi^+pn}^2(t)\left|\eta_\pi(t)\right|^2
(1-z)^{1-2\alpha_\pi(t)}
\nonumber\\ &\times&
\sigma^{\gamma^*\pi^+}_{tot}(s'=M_X^2)\,.
\label{146}
 \eeqn

The virtual photoabsorption cross section can be expressed in terms of the structure function, 
\beq
\sigma^{\gamma^*\pi^+}_{tot}(s'=M_X^2)=
\frac{4\pi^2\alpha_{em}}{Q^2}\,F_2^\pi(x_\pi,Q^2),
\label{146n}
\eeq
where
\beq
x_\pi=\frac{Q^2}{M_X^2}=\frac{x}{1-z},
\label{147n}
\eeq
and $x=Q^2/s$. 

Thus, the process of leading neutron production in DIS described by Eq.~(\ref{146}), offers a unique opportunity to measure the pion structure function at small $x$.

Experimental data are usually presented in the form of ratio of neutron production and inclusive DIS cross sections \cite{zeus-2002,zeus-2007}, which in the Born approximation can be represented as
\beqn
&&\frac{1}{\sigma_{inc}}\,\frac{d\sigma^B_{p\to n}}{dz\,dq_T^2}=
\left(\frac{\alpha_\pi^\prime}{8}\right)^2
|t|\,G_{\pi^+pn}^2(t)\left|\eta_\pi(t)\right|^2
\nonumber\\ &\times&
\frac{(1-z)^{1-2\alpha_\pi(t)}}{z}\,R_{\pi/N}(x_\pi,Q^2)\,
\frac{F_2^p(x_\pi,Q^2)}{F_2^p(x,Q^2)}\,,
\label{155n}
 \eeqn
where
\beq
R_{\pi/N}(x_\pi,Q^2)=\frac{F_2^\pi(x_\pi,Q^2)}{F_2^p(x_\pi,Q^2)}.
\label{155nn}
\eeq

The last factor in the right-hand side of Eq.~(\ref{155n}) is known and provides a sizable suppression.
Indeed, at small $x< 0.01$ the measured proton structure functions can be parametrized as 
\beq
F_2^p(x,Q^2)=c(Q^2)
\left({1\over x}\right)^{\lambda(Q^2)},
\label{1155}
\eeq
 where $\lambda(Q^2)=a\,\ln(Q^2/\Lambda^2)$ with
$a=0.048$ and $\Lambda=0.29\GeV$ \cite{h1-par}.
$F_2^p(x_\pi,Q^2)/F_2^p(x,Q^2)$ acquires a considerable suppression factor 
$(1-z)^{\lambda(Q^2)}$. For example, at $z=0.8$ and $Q^2=13\GeV^2$
(the mean value in \cite{zeus-2007}) this factor is $0.68$. This factor is the main source of $Q^2$ dependence of the fractional cross section (\ref{155n}), which turns out to be pretty weak.
For further calculations we rely on the more realistic QCD fit \cite{mstw2008}.

\subsection{What to expect?}\label{expectations}

The main unknown quantity in (\ref{155n}), which also is the main goal of experimental studies of this process, 
is the pion structure function, which enters the ratio (\ref{155nn}). Here we attempt at specifying the expected value of the ratio $R_{\pi/N}(x_\pi,Q^2)$, Eq.~(\ref{155nn}).

The hadron structure function $F_2^h(x,Q^2)$ is proportional to the total cross section of interaction of the virtual 
photon with the hadron, $F_2^h(x,Q^2)=Q^2/(4\pi^2\alpha_{em})\,\sigma^{\gamma^*p}_{tot}(x,Q^2)$.
In the target rest frame interaction with a highly virtual photon 
is a perfect counter of the number of quarks in the target.
Indeed, the interaction radius of a small ($\sim1/Q^2$) dipole  is also small ($\sim1/\ln Q^2$),
therefore interaction of the dipole simultaneously with two target valence quarks, separated by a large distance,  is suppressed. So the small dipole interacts separately with each target quark via a colorless exchange (Pomeron), i.e. the dipole-quark cross section is finite and universal, and the total dipole-hadron cross section is proportional to the number of the quarks.
One arrives at the additive quark model, which was first proposed, though ill justified,  for soft hadronic 
interactions \cite{LF}. However, a highly virtual photon interacting with a large light hadron this model should be rather accurate, so one should expect $R_{\pi/N}\equiv F_2^\pi/F_2^p=N_q^\pi/N_q^p$, where $N_q^h$ is the number of quarks in the hadron $h$. 

One can also interpret this via the QCD evolution at small $x$. The cross section of interaction of a small-size $\bar qq$ dipole with a proton is proportional to gluon density \cite{fs},
\beq
\sigma_{\bar qq}(r_T,x)=\frac{\pi^2}{3}\alpha_s(Q^2)\,xg(x,Q^2)\,r_T^2,
\label{156n}
\eeq
where $r_T\sim1/Q$ is the transverse dipole separation. 
There are many experimental evidences for existence in the proton a semi-hard scale, the mean gluon transverse momentum, of an effective gluon mass, of the order of $Q_0\sim700\MeV$ \cite{kst2,spots1,spots2}. This means that gluons are located within a small distance $\sim r_0=0.3\fm$ around the sources. Probing the proton at the semi-soft scale
$Q_0$ one resolves only the "constituent" quarks, but not their structure. At a higher scale the gluons and sea quarks are resolved as well, but the QCD evolution leaves them essentially within the same spots around the sources. Although
the radius of the spots rises with $1/x$ as $\la r^2\ra=4\alpha^\prime\ln(1/x)$, the effective slope at a hard scale is small
$\alpha^\prime\approx 0.1\GeV^{-2}$ \cite{k3p,spots2,data}, and the spots in the proton do not overlap  up to the energy of LHC \cite{k3p,spots2}.  Therefore, it is reasonable to expect that the amount of glue and sea quarks generated at small $x$ though the evolution, is proportional to the number of the quarks, which are resolved at the soft scale, and play role of the initial condition for the evolution. 
 
\subsubsection{3 valence quarks in the proton}

In the non-relativistic quark model one may expect a simple relation,
\beq
R_{\pi/N}(x_\pi,Q^2)={2\over3}, 
\label{156nn}
\eeq
provided that $x$ is sufficiently small, and $Q^2$ is large. This value was used in all previous calculations
of the cross section of leading neutron production in DIS.

It is clear, however, that the relation (\ref{156nn}) is a simplification, which misses the possibility of interaction with those constituents of the proton, which are different from just three valence quarks. Indeed, even the process under consideration is an example: as is depicted in Fig.~\ref{pion}, the virtual photon probes quarks and antiquarks in the pion cloud of the proton.

\subsubsection{A multiquark proton}

A proton experiences quantum fluctuations to the states containing more than 3 quarks. This is pretty obvious at a hard scale, since the flavor-symmetric sea of quarks and antiquarks is generated perturbatively through the QCD evolution.
Such a source of extra quarks ceases, at a soft scale, so one might think that the gluon density at small $x$ in the proton at the starting semi-hard scale is proportional to the number of valence quarks, like is assumed in Eq.~(\ref{156nn}).

There are, however, nonperturbative quantum fluctuations in the proton, which produce extra quarks at a soft scale,  also contributing to the initial conditions for the evolution. One of such mechanisms is directly related to the process under consideration. Production of leading neutrons is a part of the inclusive DIS cross section and also contributes to $F_2^p$. On the other hand, as one can see in Fig.~\ref{pion}, the small dipole $\{\bar qq\}_{\gamma^*}$ does not interact with
the 3-quark nucleons via gluonic exchanges, but interacts with the pion, i.e. with a pair of extra quarks in the proton.
Within the pion cloud model of the proton the number of quarks in the denominator of the ratio Eq.~(\ref{156nn})
should be increased:
\beq
N_q^p\Rightarrow 3+2\la n_\pi\ra,
\label{156nnn}
\eeq
where $\la n_\pi\ra$ is the mean number of pions in the proton.
The bottom bound for this correction is easy to estimate integrating the fractional cross section Eq.~(\ref{155nn}),
\beq
\frac{1}{\sigma_{inc}}\int\limits_0^1 dz\int dq_T^2\,\frac{d\sigma^B_{p\to n}}{dz\,dq_T^2} =
0.15.
\label{156m}
\eeq
Since neutral pion exchange should also provide a half of this contribution, we can estimate,
\beq
\la n_\pi\ra > 0.225.
 \label{156mm}
 \eeq
	This is the bottom bound because other final states, like baryon resonances, should also be added. This estimate is compatible with evaluations in \cite{arm-bogdan}, which ranges from $\la n_\pi\ra=0.25$ to $0.38$, dependent on the used model for the pion flux (uncorrected for absorption) and with earlier estimates in \cite{tony1,tony2,povh}.

The mean number of pions can also be evaluated basing on the observed deviation from the Gottfried sum rule \cite{gottfried} of the measured flavor asymmetry of the proton sea,
\beq
I_{AS}=\int\limits_0^1 dx\left[\bar d_p(x)-\bar u_p(x)\right]
\label{156mmm}
\eeq
The E866 experiment at Fermilab measured the value of asymmetry at $I_{AS}=0.118\pm0.012$ \cite{e866}, which results in the mean number of pions $\la n_\pi\ra=0.36$ \cite{jen-chieh}, with the usual assumption that the weight of the $|\pi\Delta\ra$ Fock state is half of that for the $|\pi N\ra$ component.
A somewhat larger values of flavor asymmetry, but with larger errors, were found in the NMC experiment,
$I_{AS}=0.148\pm 0.039$ \cite{nmc}, and HERMES, $I_{AS}=0.16\pm0.03$ \cite{hermes}, experiments.
The deduced expectations for the number of pions are $\la n_\pi\ra=0.44$ and $0.48$ respectively.

The number of quarks in the proton gets contribution not only from the flavor asymmetric, like in Eq.~(\ref{156nnn}), but also from the flavor symmetric sea. At importance of the latter indicate data \cite{e866} on $\bar d_p(x)/\bar u_p(x)$, which cannot be described by the pion cloud model \cite{tony2,jen-chieh}, and need iso-scalar contributions, like $\sigma$ and $\omega$ mesons. The analysis of this data performed in \cite{sigma} within the meson cloud model, conclude that data on $\bar d_p(x)/\bar u_p(x)$ ratio can be described with the weight factors
$\la n_\sigma\ra=0.023-0.078$ and $\la n_\omega \ra=0.063-0.671$, which may be rather large, but are quite uncertain.

Inclusion of the flavor symmetric sea into the relation (\ref{156nnn}) might considerably increase the mean number of the proton constituents at the soft scale, 
\beq
N_q^p=3+2\bigl(\la n_\pi\ra + \la n_\sigma\ra + \la n_\omega\ra\bigr).
\label{156nnnn}
\eeq
Although the contribution of the isoscalar mesons may be significant, its magnitude is model dependent and poorly known. Considering the above mentioned result of the E866 experiment, $\la n_\pi \ra=0.36$ as a lower value,
we fix the total meson contribution at $\la n_{meson} \ra=0.5$ and and use it in the following calculations.
With this value $N_q^p=4$ and instead of the simplified expectation Eq.~(\ref{156nn}), we will rely on
\beq
R_{\pi/N}(x_\pi,Q^2)={1\over2}, 
\label{final}
\eeq
Although this number has a large uncertainty, we will rely on it through all further calculations up to comparison
with DIS data for neutron production. Eventually we will check the sensitivity of data to $R_{\pi/N}$.

Notice that we have not touched so far the nominator of this ratio, the number of quarks in the pion, $N_\pi$.
Apparently, it might be also subject to corrections due to soft multiquark fluctuation, e.g. $\pi\to\pi\rho$, which is probably the strongest Fock component ($\pi\to2\pi$ is forbidden). However, pion is the Goldstone meson with an abnormally small mass, so any fluctuation is strongly suppressed by the energy denominator. In particular, the amplitude of 
the $\pi\to\pi\rho$ transition is suppressed as $m_\pi^2/(m_\rho+m_\pi)^2$. Therefore the weight factor for the 
$|\pi\rho\ra$ Fock component is so small that can be safely neglected.

\subsubsection{More uncertaities}

\begin{figure}[htb]
\vspace*{5mm}
\centerline{
  \scalebox{0.35}{\includegraphics{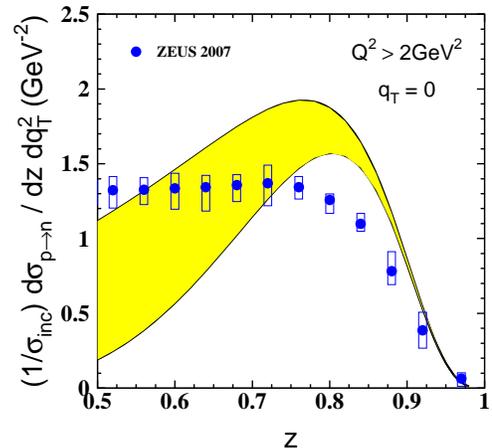}}}
\caption{\label{fig:R1-dep} (Color online) The forward fractional cross section 
of neutron production in DIS
calculated in the Born approximation, Eq.~(\ref{155n}), with $R_1^2=0$ (upper curve) and $R_1^2=2\GeV^{-2}$ (bottom curve). Data points are the results of the ZEUS experiment \cite{zeus-2007}. }
 \end{figure}

Another theoretical uncertainty in Eq.~(\ref{155n}) is related to the slope parameter $R_1^2$ of the pionic formfactor of the nucleon. 
It has to be fixed by phenomenology, but the results of model dependent analyses are quite diverse 
\cite{FMS,R1-0,ponomarev,kaidalov1,kaidalov2}    and
vary from zero to $R_1^2=2\GeV^{-2}$.
This uncertainty affects the magnitude of the fractional cross section Eq.~(\ref{155n}), especially at medium  values of  $z$. The forward cross section 
$(1/\sigma_{inc})d\sigma^B/dzdq_T^2|_{q_T=0}$
calculated in the Born approximation Eq.~(\ref{155n})  is depicted in Fig.~\ref{fig:R1-dep} by the strip between upper ($R_1^2=0$) and bottom ($R_1^2=2\GeV^{-2}$) curves. The calculations are done at $Q^2=14\GeV^2$,
which is the mean value for the DIS data \cite{zeus-2007} at $Q^2>2\GeV^2$ also depicted   Fig.~\ref{fig:R1-dep}.
 As we mentioned above, the $Q^2$ dependence of the fractional cross section is quite weak.
 
 We see that even within the uncertainty in the parameter $R_1$, the Born approximation significantly overestimates data \cite{zeus-2007} at large $z$, where the pion pole is expected to dominate. For further calculations we fix $R_1^2=0.3\GeV^{-2}$, which was chosen in \cite{k2p,kaidalov1,kaidalov2,kkmr,kmr}
 as most reliable.

\section{Absorptive corrections}\label{absorption}

Calculation of absorptive corrections, or initial/final state interactions,
is quite complicated in momentum
representation, where they require multi-loop integrations. However, these corrections factorize in impact parameters,
 \beq
f_{p\to n}(b,z)=f^B_{p\to n}(b,z)\,S(b,z,s)\,,
\label{151}
 \eeq
where $S(b,z,s)$ is the suppression factor caused by absorption. Then one can
Fourier transform the amplitude back to momentum representation,  
and the calculations are greatly simplified. So, we should first perform  Fourier
transformation of the amplitude Eq.~(\ref{100}) to impact parameter representation.

\subsection{Impact parameter representation}\label{sub-impact}

The partial Born amplitude at impact parameter $\vec b$,
corresponding to (\ref{100}), has the form,
 \beq
f^B_{p\to n}(\vec b,z)=
\bar\xi_n\left[\sigma_3\, q_L\,\theta^B_0(b,z)-
i\,\frac{\vec\sigma\cdot\vec b}{\sqrt{z}\,b}\,
\theta^B_s(b,z)\right]\xi_p,
\label{150}
 \eeq
 where
\beqn
\theta^B_0(b,z) &=& \int d^2q_T\,e^{i\vec b\vec q_T}\,
\phi^B(q_T,z)
\nonumber\\ &=&
\frac{N(z)}{1-\beta^2\epsilon^2}\,
\left[K_0(\epsilon b)-K_0(b/\beta)\right]\,;
\label{154}
 \eeqn

 \beqn
\theta^B_s(b,z) &=& {1\over b}
\int d^2q_T\,e^{i\vec b\vec q_T}\,
(\vec b\cdot\vec q)\,\phi^B(q_T,z)
\nonumber\\ &=&
\frac{N(z)}{1-\beta^2\epsilon^2}\,
\left[\epsilon\,K_1(\epsilon b)-\frac{1}{\beta}\,K_1(b/\beta)\right]\,.
\label{164}
 \eeqn
 Here
  \beq
N(z) =\frac{1}{2}\,g_{\pi^+pn}\,
z(1-z)^{\alpha^\prime_\pi(m_\pi^2+q_L^2)}
e^{-R_1^2 q_L^2}
A_{\gamma^*\pi\to X}(M_X^2)
\label{166}
\eeq
\beqn
\epsilon^2&=& z(q_L^2+m_\pi^2)\,,
\nonumber\\
\beta^2&=&{1\over z}\,\left[
R_1^2-\alpha_\pi^\prime\,\ln(1-z)\right]\,.
\label{166a}
 \eeqn

 To simplify the calculations we replaced here the Gaussian form factor,
$\exp(-\beta^2q_T^2)$, by the monopole form $1/(1+\beta^2q_T^2)$,
which is a good approximation at the small values of $q_T$ we are
interested in (both shapes are ad hoc anyway). 
At the same time we retain the Gaussian 
dependence on $q_L$, which can be rather large.

\subsection{Survival amplitude of a 
\boldmath$\{\bar qq\}_8^{\gamma^*}$-\boldmath$\{\bar qq\}_8^\pi$ dipole}

At large $z\to1$ the process under consideration is associated with
the creation of a rapidity gap, $\Delta y=|\ln(1-z)|$,
in which no particles are produced. Absorptive corrections, caused by initial 
and final state interactions of the projectile partons with the target and recoil neutron, may substantially reduce 
the probability of gap formation.
Indeed, any inelastic interaction (color exchange) of the active or spectator partons should cause intensive multiparticle production filling the gap.
Usually the corrected cross section is
calculated probabilistically, i.e. convoluting  the cross section with the survival
probability factor (see \cite{kkmr} and references therein). This
recipe may work sometimes as an approximation, but only for
$q_T$-integrated cross section. Otherwise one should rely on a
survival amplitude, rather than probability. Besides, the absorptive
corrections should be calculated differently for the spin-flip and
non-flip amplitudes (see below).

\begin{figure}[htb]
\centerline{
  \scalebox{0.27}{\includegraphics{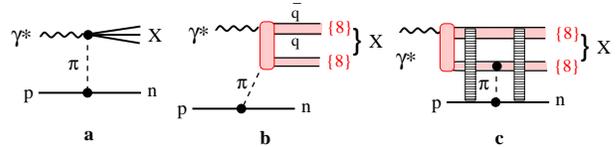}}
 }
\caption{\label{fig:fock} (Color online) 
 {\bf a:} Born graph with single pion exchange and excitation of the projectile
photon, $\gamma^*+\pi\to X$;
 {\bf b:} inelastic interaction, $\gamma^*+\pi\to X$, via
color exchange, leading to the production of two color-octet $\bar qq$ dipoles
which hadronize further to $X$;
 {\bf c:} Fock state representation of the previous mechanism. A color
octet-octet dipole as a 4-quark Fock component of the projectile photon,
$\gamma^*\to \{\bar qq\}_8-\{\bar qq\}_8$,
interacts with the target proton via
$\pi^+$ exchange. This 4-quark state may experience initial and final state
interaction via vacuum quantum number (Pomeron) exchange with the nucleons
(ladder-like strips). }
 \end{figure}

The DIS on a virtual pion shown in Fig.~\ref{pion}, i.e. the inelastic collision $\gamma^*+\pi\to X$,
can be seen as a color exchange between the colorless $\bar qq$ 
Fock component of the proton and the pion mediated by gluonic exchanges.
Nonperturbatively, e.g. in the string model, the hadron collision
looks like intersection and flip of strings.
The final state of such a collision is two color octet $\bar qq$ pairs, 
originated from the photon and pion respectively, as is depicted in Fig.~\ref{fig:fock}.
Hadronization of the
color-octet dipole,$\{\bar qq\}_8-\{\bar qq\}_8$, leads
to the production of different final states $X$.

According to Fig.~\ref{fig:fock}b the produced color octet-octet
state can experience final state interactions with the recoil
neutron. On the other hand, at high energies multiple interactions
become coherent, and one cannot specify at which point the
color-exchange interaction happens, i.e. initial and final
state interactions cannot be disentangled. In terms
of the Fock state decomposition the projectile proton fluctuates
into a 4-quark color octet-octet before the interaction with the
target. The fluctuation life-time, or coherence time,
rises with energy and at high energies considerably exceeds the
longitudinal size of target proton (see, however, more detailed discussion below). 

This leads to a different space-time picture of the process at high
energies, namely: long in advance the interaction the incoming photon fluctuates into a 4-quark state
$\gamma^*\to\{\bar qq\}_8$-$\{\bar qq\}_8$ which interacts with the target via pion exchange, as is illustrated in Fig.~\ref{fig:fock}c. 
The survival probability amplitude $S_{4q}(\vec b,\vec r,s)$ for a dipole of separation $\vec r$ colliding with a nucleon at impact parameter $\vec b$ 
can be estimated on analogy with \cite{kpss} as,
 \beq
S_{4q}(\vec b,\vec r) =1-\Im f_{4q}(\vec b,\vec r)\approx
\Bigl[1-\Im f_{\bar qq}(\vec b,\vec r)\Bigr]^2.
 \label{197}
 \eeq
 Here we rely on the large $N_c$ approximation and replaced a color octet-octet dipole by two triplet-antitriplet dipoles. 
 
 Notice that the mean
 $\{\bar qq\}_8^{\gamma^*}$-$\{\bar qq\}_8^\pi$ separation is large, and interaction of the $4q$-system with the nucleon target is soft, although the size of the $\{\bar qq\}_8^{\gamma^*}$ pair maybe as small as $1/Q$. 
Therefore, for the dipole amplitude $f_{\bar qq}(\vec b,\vec r,s)$ we employ the parametrization dependent on energy, rather than Bjorken $x$.
The $b$-integrated phenomenological dipole cross section Eq.~(\ref{360}) is parametrized in the saturated form \cite{kst2},
 \beq
\sigma_{\bar qq}(r,s)=\sigma_0(s)\left[1-e^{-r^2/R_0^2(s)}\right]\,,
\label{300}
 \eeq
 where $R_0(s)=0.88\fm\,(s_0/s)^{0.14}$; $s_0=1000\,GeV^2$;
 \beq
\sigma_0(s)=\sigma^{\pi p}_{tot}(s)
\left(1 + \frac{3\,R^2_0(s)}{8\,\la r^2_{ch}\ra_{\pi}}
\right)\,,
\label{320}
 \eeq
$\sigma^{\pi p}_{tot}(s)=23.6\mb\times(s/s_0)^{0.08}$, and the mean pion charge radius squared is $\la r^2_{ch}\ra_{\pi}=0.44\fm^2$ \cite{r-pion}.

The partial amplitude $f_{\bar qq}(\vec b,\vec r,s,\beta)$ of elastic scattering of a $\bar qq$ dipole with transverse separation $\vec r$ and fractional light-cone momenta $\beta$  (for $q$) and $1-\beta$ (for $\bar q$) on a proton at impact parameter $\vec b$
was derived  in \cite{amir1,kpss,amir2},
\begin{widetext}
\beqn
\Im f_{\bar qq}(\vec b,\vec r,s,\beta) &=&
\frac{\sigma_0(s)}{8\pi B(s)}\,
\Biggl\{\exp\left[-\frac{(\vec b-\vec r\beta)^2}{2B(s)}\right]
+\exp\left[-\frac{[\vec b+\vec r(1-\beta)]^2}{2B(s)}\right] 
\nonumber\\ &-&
2\exp\Biggl[-\frac{r^2}{R_0^2(s)}
-\frac{[\vec b+(1/2-\beta)\vec r]^2}{2B(s)}\Biggr]
\Biggr\}.
\label{340}
 \eeqn
\end{widetext}
This partial amplitude satisfies the condition,
 \beq
2\int d^2b\,{\Im}f_{\bar qq}(\vec b,\vec r,s,\beta)=
\sigma_{\bar qq}(r,s),
\label{360}
 \eeq

The fractions $\beta$ and $1-\beta$  of the light-cone momentum of the $4q$-system carries by the color octets $\{\bar qq\}_8^{\gamma^*}$ and $\{\bar qq\}_8^\pi$, which are the debris of of the photon and pion respectively, are related to $z=1-M_X^2/s$ as,
 \beq
\beta= \frac{Q^2+\la m_T^2\ra}{s(1-z)}\,,
\label{380}
 \eeq
where $m_T$ is the transverse  mass of the pion debris $\{\bar qq\}_8^\pi$, which we fix at $\la m_T^2\ra=1\GeV^2$.

The amplitude Eq.~(\ref{340}) also correctly reproduces the elastic $\pi$-$p$ slope  
$B^{\pi p}_{el}(s)$ provided that the
effective slope parameter $B(s)$ has the form \cite{kpss},
 \beq
B(s)=B^{\pi p}_{el}(s)-{1\over3}\,\la r_{ch}^2\ra_\pi
- {1\over8}\,R_0^2(s).
\label{400}
 \eeq
 We use the Regge parametrization  $B^{\pi p}_{el}(s)=B_0+2\alpha_\Pom^\prime \ln(s/\mu^2)$, with $B_0=6\GeV^{-2}$,
$\alpha_\Pom^\prime=0.25\GeV^{-2}$, and $\mu^2=1\GeV^2$.

To get the differential cross section the absorption corrected Born amplitude of neutron production, $f_{p\to n}(\vec b,\vec r,z)=f^B_{p\to n}(\vec b,z)\,S_{4q}(\vec b,\vec r,z,s)$,
 should be Fourier transferred back to the momentum representation, squared and averaged over the dipole size $r$, \beqn
&&z\,\frac{d\sigma_{p\to n}}{dz\,dq_T^2}=
\left|f_{p\to n}(q_T,z)\right|^2 =
\frac{1}{(2\pi)^4}\int d^2r\,W^2(r,M_X^2) 
\nonumber\\ &\times&
\int d^2b\, d^2b'\,e^{i\vec q_T(\vec b-\vec b')}
f_{p\to n}^\dagger (\vec b,\vec r,z)
f_{p\to n}(\vec b',\vec r,z),
\label{146nn}
 \eeqn
where $W^2(r,M_X)$ is the probability distribution of impact parameter of $\gamma^*\pi$  collision at c.m. energy $M_X$. To simplify numerical calculation we employ here the same approximation as in \cite{kpss} assuming that each amplitude can be averaged over $r$ separately, i.e. the absorption factor $S_{4q}(b,z,s)$ in (\ref{151}) is related to one in (\ref{197}) as
\beq
\bigl\la S_{4q}(\vec b,\vec r,z,s)\bigr\ra_r=\int d^2r\,W(r,Q^2,M_X^2)\,
S_{4q}(\vec b,\vec r,z,s).
\label{149n}
\eeq

To proceed further we have to specify the distribution
$W(r,M_X^2)$ over the size $r$ of the $\{\bar qq\}_8^{\gamma^*}$-$\{\bar qq\}_8^\pi$ dipole, which is 
the impact parameter of the $\gamma^*\pi$ collision at c.m. energy $M_X$.
Therefore, the $r$-distribution $W(r,M_X^2)$ is given by the partial elastic photon-pion amplitude $f_{el}^{\gamma^*\pi}(r,M_X^2)$, for which we use the normalized Gaussian $r$-dependence, 
 \beq
W(r,Q^2,M_X^2)= \frac{\exp\left[-r^2/4B^{\gamma^* \pi}_{el}(Q^2,M_X^2)\right]}
{4\pi\,B^{\gamma^* \pi}_{el}(Q^2,M_X^2)}.
\label{420}
 \eeq

With a good precision, checked numerically $\bigl\la S_{4q}(\vec b,\vec r,z,s)\bigr\ra_r\approx \bigl\la 1-\Im f_{\bar qq}(\vec b,\vec r,z,s)\bigr\ra_r^2$, where
\begin{widetext}
\beqn
\bigl\la\Im f_{\bar qq}(\vec b,\vec r,z,s)\bigr\ra_r &=& \frac{\sigma_0(s)}{8\pi}
\Biggl\{\frac{1}{{\cal B}_\beta(s,z)}
\exp\left[-\frac{b^2}
{2{\cal B}_ \beta(s,z)}\right] +
\frac{1}{{\cal B}_{1-\beta}(s,z)}
\exp\left[-\frac{b^2}
{2{\cal B}_{1-\beta}(s,z)}\right]
\nonumber\\ &-& 
\frac{2}
{{\cal B}_\xi(s,z)
\left[1+4B^{\gamma^*\pi}_{el}(M_X^2)/R_0^2(s)\right]}
\exp\left[-\frac{b^2}
{2{\cal B}_\xi(s,z)}\right]
\Biggr\}\,,
\label{440}
\eeqn
 \end{widetext}
 where
 \beq
{\cal B}_\beta(s,z) = B(s)+2\beta^2\,B^{\gamma^*\pi}_{el}(M_X^2)\,,
\label{460}
 \eeq
 
 \beqn
 {\cal B}_\xi(z,s) &=& B(s)+2\xi^2\,B^{\gamma^*\pi}_{el}(M_X^2);
 \nonumber\\
 \xi^2 &=&
\frac{(1/2-\beta)^2}
 {1+4B^{\gamma^*\pi}_{el}(M_X^2)/R_0^2(s)}
 \label{461}
 \eeqn
 
 The $Q^2$ and energy dependences of the elastic $\gamma^*\pi$ slope, $B^{\gamma^*\pi}_{el}(M_X^2,Q^2)$
are expected to be similar to what has been observed for electroproduction of different vector mesons, $\rho$, $\phi$, $J/\Psi$, in $\gamma^*p$ interactions \cite{rho-prod,phi-prod,psi-prod}. It was found that the value of the slope
saturates at $Q^2+m_V^2\gtrsim 5\GeV^2$ at the universal level, $B_{el}^{\gamma^*p}= B_0^{\gamma^*p}+2\alpha^\prime_{\Pom}\ln(W^2/\mu^2)$, with $B_0^{\gamma^*p}\approx4\GeV^{-2}$ and
$\alpha^\prime_{\Pom}\approx 0.1\GeV^{-2}$. 
Notice that the observed small value of the Pomeron trajectory slope is in a good accord with the theoretically expectation and is considerably smaller than $\alpha^\prime_{\Pom}\approx 0.25\GeV^{-2}$ observed in soft processes, which are affected by saturation of unitarity \cite{k3p, spots2}.
For the $\gamma^*\pi$ slope at high $Q^2$ we use a similar form
\beq
B_{el}^{\gamma^*\pi}(M_X^2,Q^2)= 
B_0^{\gamma^*\pi}+2\alpha^\prime_{\Pom}\ln(M_X^2/\mu^2),
\label{461a}
\eeq
with $B_0^{\gamma^*\pi}\approx B_0^{\gamma^*p}-2\GeV^{-2}$. The latter  relation is is written in analogy to the systematics of slopes observed in soft hadronic collisions. 

\begin{figure}[htb]
\centerline{
  \scalebox{0.35}{\includegraphics{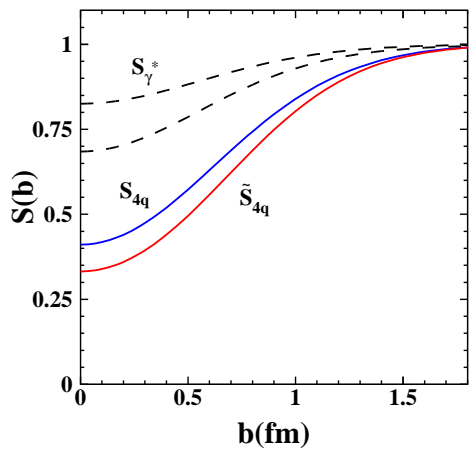}}}
 \caption{(Color online) Partial survival amplitudes $S_{4q}(b)$ (upper solid curve) calculated with Eq.~(\ref{149n}) and $\tilde S_{4q}(b)$ (bottom curve) calculated with Eq,~(\ref{490}), at c.m. energy of $\gamma^*$-proton collision $\sqrt{s}=100\GeV$ and $z=0.7$. Dashed curves  show the suppression factor $S_{\gamma^*}(b)$ calculated for virtual photons with $Q^2=3$ (bottom) and $40\GeV^2$ (upper dashed curve).}
 \label{fig:S(b)}
 \end{figure}

The result of numerical calculation of $S_{4q}(b,z)$ at $\sqrt{s}=100\GeV$, $z=0.7$ and $Q^2=13\GeV^2$
are shown as function of $b$  in Fig.~\ref{fig:S(b)} by the upper solid curve. The suppression factor hardly varies with $z$ and is slightly enhanced with energy. 
Naturally, the absorption effect is strongest for central collisions (suppression down to $40\%$), and gradually
ceases towards the periphery.

\subsection{Corrections for gluon radiation}

\begin{figure}[b]
\centerline{
  \scalebox{0.27}{\includegraphics{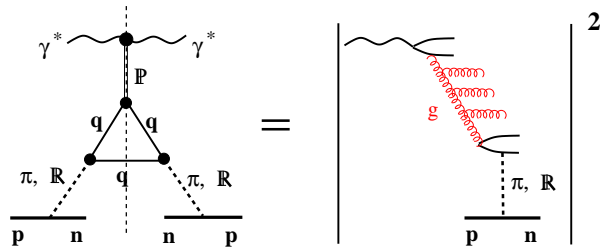}}
 } 
\caption{\label{fig:3-regge} (Color online) 
 {\sl Left:} triple-Regge graph $\pi\pi\Pom$ for leading neutron production in DIS;
 {\sl Right:} the unitarity cut of the of this graph, shown as a gluon comb. 
 }
 \end{figure}

So far our consideration was restricted to the lowest 4-quark Fock state of the photon, $\{\bar qq\}_{\gamma^*}-
\{\bar qq\}_{\pi}$, contributing to neutron production. In the triple-Regge graph shown in Fig.~\ref{fig:3-regge} (left) this Fock state would correspond to the Pomeron gluonic latter without rungs, i.e. without gluon radiation.
Gluon bremsstrahlung, is however an important process, which is responsible for the observed rise of the cross sections with energy. So higher Fock states, like one depicted in Fig.~\ref{fig:3-regge} (right), containing gluons
should be taken into account. Apparently they should lead to enhanced absorption effects.

In large-$N_c$ approximation a Fock component containing gluons can be replaced by a multi-dipole state
\cite{mueller}. The Pomeron is known to have a two-sheet topology (cylinder), which corresponds to the replacement of the 4-quark state by two $\bar qq$ dipoles as was done above. 
Every gluon radiated within one of the two Pomeron sheets creates an extra color triplet $\bar 33$ dipole.
Correspondingly, the absorption factor gains an extra suppression factor $1-\Im f_{\bar qq}(r_0)$.
Thus, a Fock state containing besides 4-quark $n$ gluons provide a modified absorption factor $\tilde S_{4q}^{(n)}$,
\beq 
S_{4q}(b)\Rightarrow \tilde S_{4q}^{(n_g)}(b)= 
S_{4q}\,\left[1-\bigl\la f_{\bar qq}(\vec b,\vec r_{gg})\bigr\ra_{r_{gg}}\right]^{n_g}.
\label{480}
\eeq

Such states containing gluons, are characterized by two different
scales, two typical dimensions \cite{spots1,spots2}.
One is the large size of the pion, $\sim1/\Lambda_{QCD}$, which dictates the mean size of the 4-quark color octet-octet dipole in (\ref{420}). Another, much smaller distance  is the mean size of a glue-glue dipole.
Analysis of data on large-mass diffraction \cite{kst2}, as well as many other experimental observables \cite{spots2}
show that this distance is quite small, $\la r_{gg}\ra\equiv r_0=0.3\fm$. Thus, averaging of the dipole amplitude,
which has the same size $r_{gg}$ as the glue-glue dipole, should be averaged as,
\beq
\bigl\la f_{\bar qq}(\vec b,\vec r_{gg})\bigr\ra =\frac{1}{\pi r_0^2}
\int d^2r_{gg}\,e^{-r_{gg}^2/r_0^2}\,
f_{\bar qq}(\vec b,\vec r_{gg})
\label{485}
\eeq

Further, we assume that the number of radiated gluons has the Poisson distribution, so we can sum up the absorption factors of the s=Fock states with different number of gluons,
\beqn
\tilde S_{4q}(b) &=& S_{4q}(b)e^{-\la n_g\ra}
\sum\limits_{n_g=0}\frac{\la n_g\ra^{n_g}}{n_g!}
\left[1-\bigl\la f_{\bar qq}(\vec b,\vec r_{gg})\bigr\ra\right]^{n_g}
\nonumber\\ &=&
S_{4q}(b)e^{-\la n_g\ra\la f_{\bar qq}(b)\ra}.
\label{490}
\eeqn

The mean number of radiated gluons $\la n_g\ra$ can be estimated looking at the $x$-dependence of the DIS cross section. In the leading-log approximation integration over rapidity of each gluon results in a factor $\ln(1/x)$.
Summing over gluon number one gets for the total $\gamma^*$-proton cross section,
\beq
\sigma_{tot}^{\gamma^*p}\propto
\sum\limits_{n_g=0}
\frac{\bigl[g\ln(1/x)\bigr]^{n_g}}{n_g!}=
\left({1\over x}\right)^g,
\label{492}
\eeq
where $g$ includes the coupling and other factors acquired due to radiation of each gluon.

Thus, according to Eq.~(\ref{1155}) the mean number of radiated gluon reads,
\beq
\la n_g\ra=\lambda(Q^2)\ln(1/x_\pi),
\label{494}
\eeq
where $\lambda(Q^2)$ is defined in (\ref{1155}).

Now we are in a position to calculate the modifies absorption factor Eq.~(\ref{490}).
The result is plotted by the bottom solid curve in Fig.~\ref{fig:S(b)} in comparison
with the uncorrected survival amplitude $S_{4q}$. Although the modified suppression is
stronger, as was anticipated, the difference is rathe small. This is a result of smallness of $r_0$.

\subsection{Coherence lengths for the photon Fock states}

We have assumed so far that all the Fock components of the photon considered above have the lifetime,
or coherence length, much longer than the dimension of the target. This is certainly true for the simplest Fock state
$\gamma^*\to\bar qq$, which has a long coherence time, called Ioffe time (see more accurate evaluation in \cite{krt-dis}), 
\beq
l_c^{\bar qq}=\frac{1}{2xm_N},
\label{495}
\eeq
where $x\sim 10^{-3}$ in the kinematics of HERA.

However for the 4-quark Fock states this is not obvious. The coherence length is given by
\beq
l_c^{4q}=\frac{1}{q_L}=\frac{\sqrt{z}}{(1-z)m_N}.
\label{496}
\eeq
This coherence time becomes very long for large $z\to1$

Since the target nucleon size is $r_N\sim 1\fm$, only at $z\gsim 0.8$ the coherence length Eq.~(\ref{496})
is sufficiently long to rely on the above evaluations of the absorptive effects.

In another limit of a very short coherence length $l^{4q}_c\ll r_N$ not only initial, but also final state interactions 
of the 4-quark state are impossible, because even the the interaction time and time scale of creation of this state exceed the nucleon size. In this case the absorptive corrections are generated only by the interaction of  long-living $\bar qq$ fluctuations of the photon. In this case the absorption factor has the form,
\beq
S_{\gamma^*}(b)=1-\bigl\la\Im f_{\bar qq}(\vec b,\vec r,\alpha)\bigr\ra_{r, \alpha},
\label{497}
\eeq
where the partial elastic amplitude $\Im f_{\bar qq}(\vec b,\vec r, \alpha)$ is given by Eq.~(\ref{340}).
The averaging over the transverse dipole separation $\vec r$, and the fractional light-cone momentum of the quark,
$\alpha $, is  done as follows,
\begin{widetext}
\beq
\bigl\la\Im f_{\bar qq}(\vec b,\vec r, \alpha)\bigr\ra_{r, \alpha} =
\left[\int\limits_0^1 d \alpha\int d^2r \sigma_{\bar qq}(r,x_\pi)\,\left|\Psi_{\bar qq}(r, \alpha,Q^2)\right|^2\right]^{-1}
\int\limits_0^1 d \alpha\int d^2r \sigma_{\bar qq}(r,x_\pi)\left|\Psi_{\bar qq}(r, \alpha,Q^2)\right|^2
\Im f_{\bar qq}(\vec b,\vec r, \alpha),
\label{498}
\eeq
\end{widetext}
where the weight factor $\left|\Psi_{\bar qq}\right|^2=\left|\Psi_{\bar qq}^T\right|^2+
\left|\Psi_{\bar qq}^L\right|^2$ contains the standard photon distribution functions \cite{bjorken1,bjorken2}. Notice the importance of the factor $\sigma_{\bar qq}(r,x_\pi)$, which comes from the Born amplitude of neutron production. Without this factor
the result of (\ref{498}) would be zero \cite{krt-dis}, because the normalization of the distribution function of transversely polarized photons, $\Psi_{\bar qq}^T$ is ultraviolet divergent. This divergency corresponds to ultra-heavy $\bar qq$ fluctuations, which dominate in a transversely polarized photon in vacuum. However, such fluctuation are "sterile", i.e. cannot interact (color transparency), while the process under consideration contains at least one dipole interaction.

Examples of the results for $S_{\gamma^*}(b)$ calculated at $Q^2=3$ and $40\GeV^2$ are plotted in Fig.~\ref{fig:S(b)} by dashed curves. Apparently, the $Q^2$ dependence of the absorption factor is rather weak. This was anticipated, because
in (\ref{498}) we average the dipole amplitude squared, similar to diffraction of nuclear shadowing. The integral
over $\alpha$ turns out to be dominated by the aligned-jet configurations \cite{aligned-jet}, i.e. by the endpoint  behavior of the distribution functions, $\alpha \lsim m_q^2/Q^2$ \cite{k-povh}. The corresponding $\bar qq$ transverse separation becomes rather large, $\sim1/m_q$, and independent of $Q^2$, $\la r^2\ra\sim Q^2\alpha(1-\alpha)+m_q^2$. We fixed the effective quark mass, which is in fact the infrared cutoff, at $m_q=0.15\GeV$ adjusted to data on nuclear shadowing \cite{krt-dis}.

\begin{figure}[htb]
\centerline{
  \scalebox{0.35}{\includegraphics{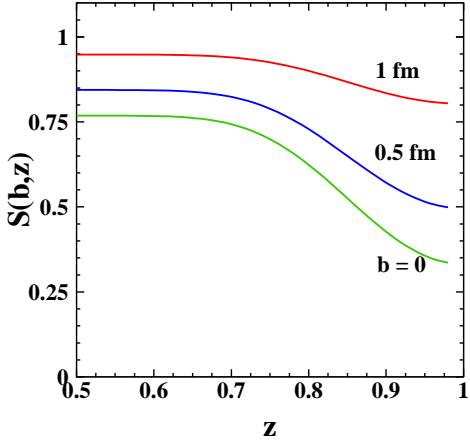}}}
 \caption{(Color online) The absorption factor $S(b,z)$, Eq.~(\ref{499}) as function of $z$ at fixed impact parameters $b=0,\ 0.5$ and $1\fm$.}
 \label{fig:S(z)}
 \end{figure}

Thus, we know the absorption corrections in two limiting regimes: (i) $l_c^{4q}\gg r_N$, in this case the suppression factor $\tilde S_{4q}(b)$ is given by Eq.~(\ref{490}) and depicted by solid curves in Fig.~\ref{fig:S(b)}; (ii) $l_c^{4q}\ll r_N$, in this case the absorption factor $S_{\gamma^*}(b)$ is given by Eq.~(\ref{497}) and is shown by solid curves in Fig.~\ref{fig:S(b)}. In order to interpolate between these limiting regimes we employ the following 
simple procedure,
\beq
S(b)=\tilde S_{4q}(b)\,F_N(q_L)+S_{\gamma^*}(b)\bigl[1-F_N(q_L)\bigr],
\label{499}
\eeq
where the transition formfactor is chosen in the dipole form, 
 $F_N(q_L)=\bigl(1+q_L^2 L^2 \bigr)^{-1}=\left[1+\bigl(L/l_c^{4q}\bigr)^2\right]^{-1}$.
The parameter $L$ characterizes the dimension of the target nucleon, so it should be of the order of $1\fm$, and we fix it at this value, $L=1\fm$, for further calculations. However, this parameter can be varied within a reasonable range. The $z$-dependence of the suppression factor Eq.~(\ref{499}) is plotted in Fig.~\ref{fig:S(z)} for few values of $b$.

Notice that although the recipe Eq.~(\ref{499})  is partially ad hoc, it interpolates between the two known limiting regimes of very long, $l_c^{4q}\gg r_N$ ($z\gtrsim0.8$), and very short, $l_c^{4q}\ll r_N$ ($z\sim 0.5$) coherence length. As far as these two regimes are predicted, the interpolation procedure should not affect the results significantly.

\subsection{Cross section corrected for absorption}\label{xsection}

As soon as the absorption factor Eq.~(\ref{499}) is known, we can perform the inverse Fourier transformation to momentum representation,
\beqn
f_{p\to n}(q_T,z)&=&
\frac{1}{(2\pi)^2}
\int d^2b\,e^{i\vec q_T\cdot\vec b}
\label{148n}\\ &\times&
f^B_{p\to n}(\vec b,z)\,S(b,z,Q^2,s)
\nonumber
 \eeqn
where the Born amplitude in impact parameters, $f^B_{p\to n}(\vec b,z)$ is given by (\ref{150}). So the absorption corrected partial spin amplitudes read,
 \beq
\theta_{0,s}(b,z)=
\theta^B_{0,s}(b,z)\,
S(b,z).
\label{500}
 \eeq

Then the Fourier transformed amplitude reads,
 \beq
A_{p\to n}(\vec q_T,z)=
\bar\xi_n\left[\sigma_3 q_L\,\phi_0(q_T,z)+
\vec\sigma\vec q_T\frac{\phi_s(q_T,z)}{\sqrt{z}}\right]\xi_p,
\label{520}
 \eeq
 where according to (\ref{154}), (\ref{164}) and (\ref{166}),
 \beqn
\phi_0(q_T,z)&=&\frac{N(z)}{2\pi(1-\beta^2\epsilon^2)}
\int\limits_0^\infty db\,b\,J_0(bq_T)\,
S(b,z)
\nonumber\\ &\times&
\left[K_0(\epsilon b)-K_0\left({b\over\beta}\right)\right]
\,;
\label{540}
 \eeqn

 \beqn
q_T\,\phi_s(q_T,z)&=&\frac{N(z)}{2\pi(1-\beta^2\epsilon^2)}
\int\limits_0^\infty db\,b\,J_1(bq_T)\,
S(b,z)
\nonumber\\ &\times&
\left[\epsilon\, K_1(\epsilon b)-
{1\over\beta}\,K_1\left({b\over\beta}\right)\right]
\,.
\label{560}
 \eeqn

Now we can calculate the differential cross
section of inclusive production of neutrons corrected for absorption,
\beq
z\,\frac{d\sigma_{p\to n}}{dz\,dq_T^2}=
\sigma_0(z,q_T) + \sigma_s(z,q_T)\,,
\label{580}
 \eeq
 where
 \beqn
\sigma_0(z,q_T)&=&
\frac{q_L^2}{s}\,
\left|\phi_0(q_T,z)\right|^2
\label{590}\\
\sigma_s(z,q_T)&=&
\frac{q_T^2}{zs}\,
\left|\phi_s(q_T,z)\right|^2\,.
\label{600}
 \eeqn
 
\begin{figure}[htb]
\centerline{
  \scalebox{0.35}{\includegraphics{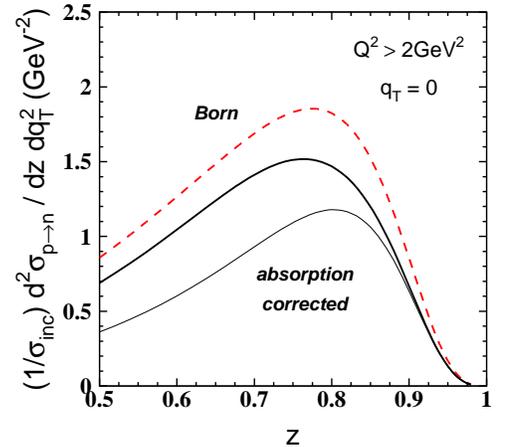}}}
 \caption{(Color online)  Forward fractional cross section of neutron production in DIS calculated without absorption corrections (dashed curve); absorption suppressed by factor $\tilde S_{4q}(b)$ Eq.~(\ref{490}) (thin solid curve);
 and with the $z$-dependent absorption factor $S(b,z)$ Eq.~(\ref{499}) (thick solid curve). 
 }
 \label{fig:abs-corr}
 \end{figure}

The effects of absorptive corrections are illustrated in Fig.~\ref{fig:abs-corr}
The effect of suppression caused by the absorption factor $\tilde S_{4q}(b)$ defined in (\ref{590})
is demonstrated by comparison of the fractional cross section of forward neutron production (dashed curve) with
absorption suppressed result plotted by the {\sl thin} solid curve. We observe a rather strong effect, absorption reduces the cross section by nearly factor 2. Inclusion of the coherence time effect results in 
a $z$-dependent absorption factor $S(b,z)$ defined in (\ref{499}) and illustrated in Fig.~\ref{fig:S(z)}.
This final absorption corrected cross section is plotted in FIg.~\ref{fig:abs-corr} by {\sl thick} solid curve.
We postpone comparison with data, because several more mechanisms of neutron production are to be added.

\section{Other Reggeons}\label{reggeons}

Besides pion exchange, other iso-vector Reggeons contribute to neutron production.
Those are subdivided to natural parity Reggeons ($\rho,\ a_2$), which have high intercepts $\alpha_R(0)\approx 1/2$, and unnatural parity ones ($a_1$, $\pi'$, etc.) with lower intercepts

\subsection{Natural parity Reggeons}

The leading Reggeons contributing to neutron production are $\rho$ and $a_2$. At large invariant masses $M_x$ they do not interfere with the pion exchange and with each other. Indeed, summing over final states at fixed $M_X$ one gets the imaginary parts of the amplitudes for the processes, $\gamma^*+\pi\to\gamma^*+\rho$, or $\gamma^*+\rho\to\gamma^*+a_2$, which  
are suppressed by a power of $1/M_X$. 

These amplitude are known to be dominated by their spin-flip part \cite{kane,irving}, so we neglect the non-flip term in what follows. In the Born approximation the leading  Reggeons contribute to the cross section as,
\beqn
&&\frac{1}{\sigma_{inc}}\left(\frac{d\sigma^B_{p\to n}}{dz\,dq_T^2}\right)_{\!\!R}=
q_T^2\left(\frac{\alpha_R^\prime}{8}\right)^2
\frac{(1-z)^{1-2\alpha_R(t)}}{z^2}\,
\nonumber\\ &\times&
G_{R^+pn}^2(t)
\left|\eta_R(t)\right|^2
\frac{F_2^R(x_\pi,Q^2)}{F_2^p(x,Q^2)}\,.
\label{620}
 \eeqn
We consider two leading exchange degenerate Reggeons $\rho$ and $a_2$. The signature factor of the latter 
$\eta_{a_2}(t)=-i-\cot\left[{\pi\over2}\alpha_{a_2}(t)\right]$ diverges at the so called nonsense wrong signature point, $t_0=-\alpha_{a_2}(0)/\alpha_{a_2}^\prime\approx-0.6\GeV^2$. In order to kill this unphysical  pole
the residue function of the $a_2$ Reggeon must have a zero at this point, i.e. a factor $(1-t/t_0)$. According to exchange degeneracy the residue function of the $\rho$-Reggeon should also contain this factor, which is not compensated  by any pole at $t=t_0$. This is confirmed by data on differential cross section of reaction $\pi^-p\to\pi^0n$, which indeed has a dip at $t=t_0$ \cite{irving}.

In this circumstances the $t$-dependences of $\rho$ and $a_2$ exchange amplitudes are rather uncertain.
Since we are focused on small-$t$ region, the most reasonable solution seems to be to fix the signature factors of both Reggeons at $\eta_R(0)$. Moreover, basing on the exchange degeneracy of $\rho$ and $a_2$ Reggeons, we fix 
$\alpha_\rho(t)=\alpha_{a_2}(t)$ and 
$G_{\rho^+pn}(t)=G_{a_2^+pn}(t)$.

The contribution to the cross section of the spin-flip Reggeon amplitude can be described by the term similar to Eq.~(\ref{600}), properly modified,
\beq
\left(\frac{d\sigma_{p\to n}}{dz\,dq_T^2}\right)_{\!\!R}=
\frac{2}{z}\sigma^\rho_s(z,q_T)=
\frac{2q_T^2}{z^2s}
\left|\phi^\rho_s(q_T,z)\right|^2\,,
\label{630}
 \eeq
where $\phi^\rho_s(q_T,z)$, compared to Eq.~(\ref{560}), contains the imaginary part neglected for pions,
and several modifications, 
\beq
\phi^\rho_s(q_T,z)=
\frac{N_\rho(z)}{2\pi q_T\beta_\rho^3}
\int\limits_0^\infty db\,b\,J_1(bq_T)
K_1(b/\beta_\rho)S(b,z).
\label{640}
\eeq
 The further notations are,
  \beqn
N_\rho(z) &=&\frac{\pi\,\alpha_\rho^\prime}{4}\,g_{\rho^+pn}\,\eta_\rho(0)
z(1-z)^{-\alpha_\rho(0)+\alpha^\prime_\rho\,q_L^2}
\nonumber\\&\times&
e^{-R_\rho^2 q_L^2}
A_{\gamma^* \rho\to X}(M_X^2)
\label{660} \\
\beta_\rho^2&=&{1\over z}\,\left[
R_\rho^2-\alpha_\rho^\prime\,\ln(1-z)\right]\,.
\label{680}
 \eeqn
 
 The absorptive corrections for the Reggeons are calculated with the same suppression factor $S(b)$ in (\ref{640}). Its effect maybe somewhat stronger for Reggeons than for pions, since the interaction in this case is more central.
 
 For the vertex function $G_{\rho NN}(t)=g_{\rho NN}\exp(R_{\rho}^2t)$ we rely on the phenomenological global Regge analysis \cite{irving} of high-energy hadronic data, which results in $g_{\rho NN}= 0.5\, g_{\pi NN}$, and $R^2_\rho=1\GeV^{-2}$.

Notice that $N_\rho(z)$ contains an additional $z$-dependence, a factor $\sim1/\sqrt{1-z}$, compared to 
the pion exchange, Eq.~(\ref{166}). 
This mean that at $z\to1$ the $\rho/a_2$ Reggeon contribution should dominate over the pion exchange \cite{k2p},
although at which $z$ this happens depends on the couplings strength.

\subsection{\boldmath$a_1$-like exchanges}

The study of spin effect in leading neutron production performed recently \cite{kpss-spin} revealed important role of the axial-vector Reggeons, like $a_1$ meson. Interference of the related non-flip spin amplitude with
spin-flip pion exchange well explained data on transverse single-spin asymmetry of leading neutrons produced in $pp$ collisions.

If fact, the situation with axial vector mesons is more complicated. Assuming vector meson dominance in the axial current, in analogy with the vector current, one arrives at a dramatic contradiction of the Adler relation for diffractive neutrino-production of pins with data,
the effect called Piketti-Stodolsky puzzle \cite{pik-stod}. It was  proposed in \cite{marage} that the source of the problem is the assumed axial-vector dominance, while in reality the $a_1$ pole is a very weak singularity,
and the main contribution to the dispersion relation for the axial current comes from the $\rho$-$\pi$ cut.
Indeed a detailed analysis of data on diffractive dissociation $\pi\to\rho\pi$ performed in \cite{kpss-spin} shows that
the invariant mass distribution of the produced in the $1^+S$ wave $\rho$-$\pi$ forms a pronounced narrow peak at a mass $M_{\tilde a_1}=1.12\GeV$ close to the $a_1$ mass. In many instances one can tread such a $\rho$-$\pi$ Regge cut as an effective $\tilde a_1$-pole \cite{belkov,marage,pcac,kpss-spin}.
Its contribution to the spin nin-flip part of the Born amplitude Eq.~(\ref{100}) reads,
 \beqn
\bigl[\phi^{\tilde a_1}_0(q_T,z)\bigr]^B&=&\frac{\alpha_{\tilde a_1}^\prime}{8}\,
G_{\tilde a_1pn}(t)\,\eta_{\tilde a_1}(t)
(1-z)^{-\alpha_{\tilde a_1}(t)}
\nonumber\\ &\times&
A_{\gamma^* \tilde a_1\to X}(M_X^2)\,,
\label{700}
 \eeqn
where
\beq
\eta_{\tilde a_1}(t)=-i-tg\!\left[\frac{\pi\alpha_{\tilde a_1}(t)}{2}\right].
\label{720}
\eeq
The Regge trajectory of the $\rho$-$\pi$ cut has the form,
\beq
\alpha_{\tilde a_1}(t)=\alpha_{\pi\rho}(t)=\alpha_\pi(0)+\alpha_\rho(0)-1+
\frac{\alpha_\pi^\prime\alpha_\rho^\prime}
{\alpha_\pi^\prime+\alpha_\rho^\prime}\,t ,
 \label{780} 
 \eeq 
so $\alpha_{\tilde a_1}(0)=-0.5$; $\alpha_{\tilde a_1}^\prime=0.45\GeV^{-2}$.

The $\tilde a_1NN$ vertex is parametrized as $G_{\tilde a_1pn}(t)=g_{\tilde a_1pn}\exp(R_{\tilde a_1}^2t)$.
The ${\tilde a_1}NN$ coupling was evaluated in \cite{kpss-spin} basing on PCAC and the 
second Weinberg sum rule, in which the spectral functions of the vector and axial
currents are represented by the $\rho$ and the effective ${\tilde a_1}$ poles respectively.
This leads to the following relations between the couplings,
\beq
\frac{g_{\tilde a_1 NN}}{g_{\pi NN}}=
\frac{m_{\tilde a_1}^2\,f_\pi}{2m_N\,f_\rho}\approx 0.5,
\label{740}
\eeq
where $f_\pi=0.93m_\pi$ is the pion decay coupling;
$f_\rho=\sqrt{2}m_\rho^2/\gamma_\rho$, 
and $\gamma_\rho$ is the universal coupling ($\rho NN$,
$\rho\pi\pi$, etc), $\gamma_\rho^2/4\pi=2.4$.

Applying to the Born amplitude Eq.~(\ref{700}) the procedure of correcting for absorption,
developed above, we Fourier transform the Born amplitude to impact parameters, introduce the absorption factor $S(b)$, the transform the result back to momentum representation and get,
\beq
\phi^{\tilde a_1}_0(q_T,z)=
\frac{N_{\tilde a_1}(z)}{2\pi \beta_{\tilde a_1}^2}
\int\limits_0^\infty db\,b\,J_0(bq_T)
K_0(b/\beta_{\tilde a_1})S(b,z),
\label{820}
\eeq
where
  \beqn
N_{\tilde a_1}(z) &=&\frac{\pi\,\alpha_{\tilde a_1}^\prime}{4}\,g_{{\tilde a_1}pn}\,\eta_{\tilde a_1}(0)
z(1-z)^{-\alpha_{\tilde a_1}(0)+\alpha^\prime_{\tilde a_1}q_L^2}
\nonumber\\&\times&
e^{-R_{\tilde a_1}^2 q_L^2}
A_{\gamma^* {\tilde a_1}\to X}(M_X^2)
\label{660a} \\
\beta_{\tilde a_1}^2&=&{1\over z}\,\left[
R_{\tilde a_1}^2-\alpha_{\tilde a_1}^\prime\,\ln(1-z)\right]\,.
\label{840}
 \eeqn
 For the sake of simplicity we assume that $A_{\gamma^* {\tilde a_1}\to X}(M_X^2)$ is equal to the same amplitude on pion, $\rho$, or $a_2$ targets.
  
Besides, the interference between spin non-flip amplitudes with pion and $a$ exchanges also contributes. Altogether 
the corresponding part of the cross section reads,
 \beqn
\sigma_0(z,q_T)&=&
\frac{q_L^2}{s}\,\Bigl[
\left|\phi_0^\pi(q_T,z)\right|^2 +
\xi^2\left|\phi_0^{\tilde a_1}(q_T,z)\right|^2 \nonumber\\ &+&
2\xi\Re \phi_0^{\pi {\tilde a_1}}(q_T,z)\Bigr],
\label{860}
\eeqn
where $\xi=2m_N/\sqrt{|t|}\approx 2/(1-z)$ is a factor related to the spin structure of the axial-vector vertex \cite{kpss-spin}.

 In the interference term one needs to know the off-diagonal diffractive amplitude,
\beq
\sum\limits_{X}
A_{\gamma^* {\tilde a_1}\to X}^\dagger A_{\gamma^* \pi\to X}=
M_X^2\,\Im A_{\gamma^* \pi\to \gamma^*{\tilde a_1}}(M_X^2,p_T=0)
\label{880}
\eeq
The amplitude of the process $\gamma^*\pi\to\gamma^*{\tilde a_1}$ can be related to the reaction $\pi p\to {\tilde a_1}p$ relying on Regge factorization, and dominance of the diffractive excitation $\pi\to \rho\pi$ (see above).
The amplitude Eq.~(\ref{880}) 
is suppressed compared with $\gamma^*\pi$ elastic scattering by the factor $\Omega$ defined as,
\beqn
\Omega^2&\equiv&\frac{d\sigma(\gamma^*\pi\to\gamma^* {\tilde a_1})/dp_T^2}
{d\sigma(\gamma^*\pi\to\gamma^*\pi)/dp_T^2}\Bigr|_{p_T=0}
\nonumber \\ &\approx&
\frac{d\sigma(\pi p\to\pi\rho p)/dp_T^2}
{d\sigma(\pi p\to\pi p)/dp_T^2}\Bigr|_{p_T=0}.
\label{900}
\eeqn
According to \cite{kpss-spin} $L$ is energy independent and can be evaluated at $M_X^2=150\GeV^2$ where data for diffractive $\pi p$ interactions are available,
$d\sigma(\pi p\to\tilde a_1 p)/dp_T^2\bigr|_{p_T=0}=1.67\mb/\!\GeV^2$. Then,
\beq
\Omega=\frac{\sqrt{16\pi\times1.67\mb/\GeV^2}}{\sigma^{\pi p}_{tot}}=
0.29.
\label{920}
\eeq

 Thus, the third term in (\ref{860}) can be presented as,
 \beqn
 2\Re \phi_0^{\pi {\tilde a_1}}(q_T,z)&=& 
 - 2\Omega\,\frac{\tan\left[\pi\alpha_{\tilde a_1}(-q_L^2)/2\right]
}{|\eta_{\tilde a_1}(0)|}\,
  \nonumber\\ &\times&
\sqrt{\left|\phi_0^\pi(q_T,z)\right|^2
 \left|\phi_0^{\tilde a_1}(q_T,z)\right|^2},
  \label{940}
 \eeqn
where $\alpha_{\tilde a_1}(t)$ is given by Eq.~(\ref{780}) and we neglected the small imaginary part of the pion signature function Eq.~(\ref{140}).

Notice that the contribution of the $\tilde a_1$ exchange to neutron production has been well tested. Analogous calculations \cite{kpss-spin} of the imaginary part of the $\pi-{\tilde a_1}$ interference led to a very good agreement with data on azimuthal asymmetry of neutrons produced by polarized protons.

 \section{Numerical results}
\label{results} 
 \subsection{\boldmath$z$-dependence}
 
\begin{figure}[htb]
\centerline{
  \scalebox{0.35}{\includegraphics{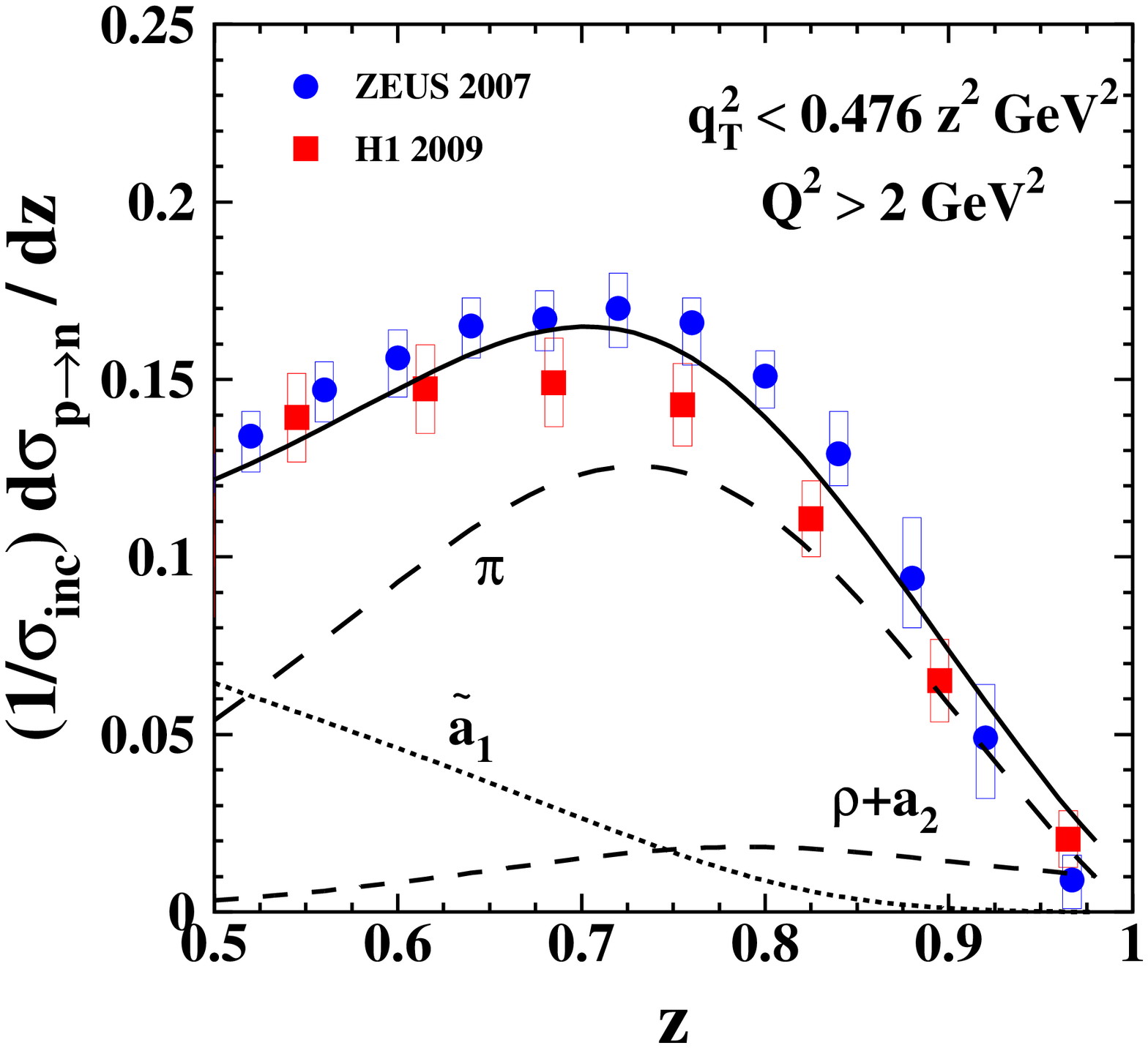}}}
  \vspace{5mm}
\centerline{
  \scalebox{0.35}{\includegraphics{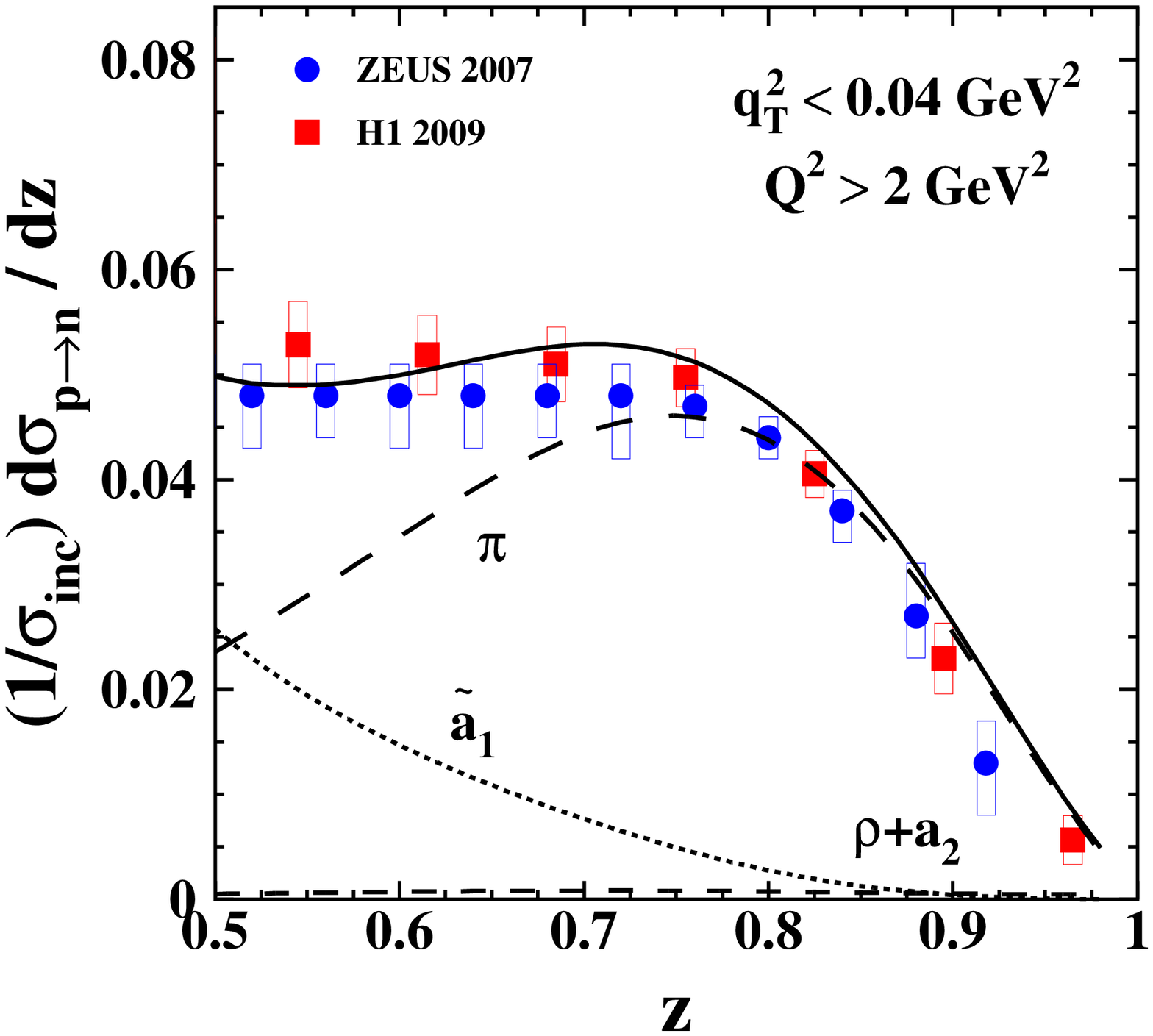}}
}
 \caption{Fractional differential cross section of neutron production,
integrated up to $q_T^{max}=0.2\GeV$ (upper panel) and 
up to $q_T^{max}=z\times0.69\GeV$
(bottom panel).  ZEUS data \cite{zeus-2007} are shown by round points.
Data from the H1 experiment \cite{h1} depicted by squares are normalized as described in the text.
In all sets of data the error are dominated by the systematic uncertainties. The pion pole contribution, calculated including absorptioncorrections is plotted by long-dashed curve (compare with Fig.~\ref{fig:abs-corr}). The contribution of $\rho$ and $a_2$ Reggeons and the effective $\tilde a_1$ pole are shown by short-dashed and dotted curves respectively. The sum of all these terms is presented by solid curve.
}
 \label{fig:qt-integrated}
 \end{figure}

 Two experiments at HERA, ZEUS \cite{zeus-2002,zeus-2007} and H1 \cite{h1}  have studied leading neutron production in DIS. Their results are available for comparison with the same kinematics, integrated over $q_T$ 
within a fixed polar angle,
$q_T<z\times 0.69\GeV$,
and up to a fixed maximum, $q_T<0.2\GeV$. These data are depicted by round points in Fig.~\ref{fig:qt-integrated} in the upper and bottom panels respectively.
Analogous results of the H1 experiment are published in \cite{h1}, in the form of absolute values of the cross section. To compare with the ZEUS data and our calculations we normalized the H1 data dividing by inclusive DIS cross section $\sigma_{inc}=112\,{\rm pb}$ \cite{armen}. The results are plotted in Fig.~\ref{fig:qt-integrated}
by squares.  Data for the fractional cross section of both experiments agree with each other  if integrate in the large interval $q_T<z\times 0.69\GeV$. However at small $q_T<0.2\GeV$ the H1 cross section considerably exceeds the one measured by ZEUS at $z\lesssim 0.8$. This indicates at a significantly different $q_T$-slopes of the differential cross section measured in these two experiments.
 
The absorption corrected pion Regge pole contribution, which was shown by thick solid curve in Fig.~\ref{fig:abs-corr}, is depicted here by long-dashed curves. 
As was anticipated, the contribution of the effective $\tilde a_1$-Reggeon plotted by dotted curve, is vanishing at large $z$ because of the low Regge intercept. On the contrary, the $\rho$ and $a_2$ Reggeons are increasingly important towards large $z$, and even dominate at $z\to1$.

\subsection{\boldmath$Q^2$-dependence}
\begin{figure}[htb]
\centerline{
  \scalebox{0.35}{\includegraphics{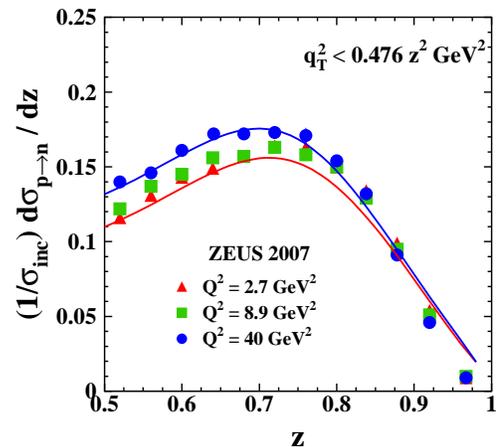}}}
 \caption{(Color online) $Q^2$ dependence of the fractional DIS cross section of neutron production,
integrated up to $q_T^{max}=z\times0.69\GeV$.  ZEUS data \cite{zeus-2007} for $Q^2=2.7,\ 8.9$ and $40\GeV^2$ are shown by triangles, squares and round points respectively. The upper and bottom curves are calculated including all considered mechanisms at $Q^2=40$ and $2.7\GeV^2$.}
 \label{fig:Q2-dep}
 \end{figure}

The full collections of ZEUS data named DIS with $Q^2>2\GeV^2$ can be binned in order to trace the $Q^2$ dependence of the cross section. The result of such a binning is presented in Fig.~\ref{fig:Q2-dep}.     
It demonstrate a clear increase of the fractional cross section with $Q^2$, especially at $z<0.8$.
The variation with $Q^2$ of the calculated cross section originates solely from the absorption factor $S_{\gamma^*}$ (see Fig.~\ref{fig:S(b)}) which is increasingly important towards small $z$ according to Eq.~(\ref{499}).
This naturally explains the observed trend of a weakened  $Q^2$ dependence at large $z$.

\subsection{\boldmath$q_T$-dependence}

\begin{figure}[htb]
\centerline{
  \scalebox{0.45}{\includegraphics{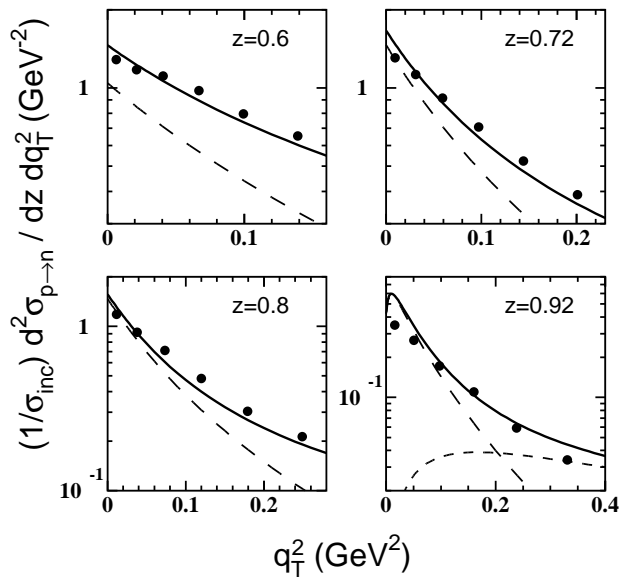}}
}
 \caption{Fractional differential cross section of leading neutron production as function of neutron transverse momentum at  several values of $z$ (solid curves).  The contributions of pure pion exchange and of the iso-vector Reggeons, other than pion, are depicted by long and short-dashed curves respectively.
 ZEUS dara are plotted by round points. The large systematic errors (see \cite{zeus-2007}) are not shown.}
 \label{fig:qt-dep}
 \end{figure}

We also compare the $q_T$ dependence of the fractional differential cross section with few samples of ZEUS data 
presented in Fig.~\ref{fig:qt-dep}. 
Our results are shown by solid curves, which sum up the pion (long-dashed curves) and Reggeon (short-dashed curves) contributions. 
Apparently, the role of Reggeons increases with $z$, especially at large $q_T$,
where they significantly diminish the slope compared with the net pion contribution. 
Correspondingly, the calculated $q_T$ distribution acquires a bent shape, while  the ZEUS data seem to prefer the Gaussian $q_T$-shape.

According to Eq.~(\ref{130}) at large $z$ and small $q_T$ one approaches the pion pole, $1/(t+m_\pi^2)$.
Therefore, the $q_T$-distribution of neutron production at large $z$ is expected to be rather steep, since is controlled by  the small pion mass. However, the relative contribution of natural parity Reggeons rises, and eventually, they take over at $z\to1$. Therefore, one could expect a sudden drop of the slope of the $q_T$-distribution.
 
Few examples of the $q_T$-dependence of the differential cross section plotted by solid curves, are compared with ZEUS data in Fig.~\ref{fig:qt-dep}. The contributions given by Eq.~(\ref{580}) and of Reggeons are plotted by long- and short-dashed curves respectively. Apparently, the Reggeons are important to achieve agreement with data.

\begin{figure}[!ht]
\centerline{
  \scalebox{0.35}{\includegraphics{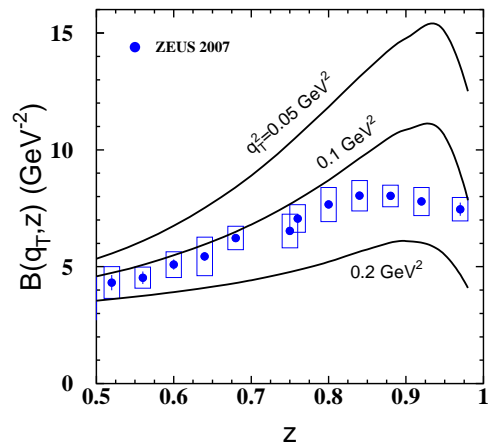}}
}
 \caption{(Color online) ZEUS data \cite{zeus-2007} for the $q_T$-slope of the differential DIS cross section of leading neutron production as function of $z$. The solid curves present theoretical predictions calculated with Eq.~(\ref{960}) at fixed $q_T^2=0.5,\ 0.1$ and $0.2\GeV^2$.}
 \label{fig:slope}
 \end{figure}

Data of ZEUS are also presented in \cite{zeus-2007} as the slope of the $q_T$ distribution versus $z$, as is shown in Fig.~\ref{fig:slope}.
Theoretically, the slope is ill defined, if one does not specify in which interval of $q_T$ it was measured. 
The local slope, defined as
\beq
B(q_T)=\frac{\partial}{\partial q_T^2}\,
\ln\left[\frac{d\sigma_{p\to n}}{dz dq_T^2}\right],
\label{960}
\eeq
apparently may vary with $q_T$. Although the results of the ZEUS experiment agree with $q_T$-independent slope (within large systematic errors), this is certainly not the case in the theory. For this reason we do not perform an explicit comparison of our results with data, but present in Fig.~\ref{fig:slope} the $z$-dependent slope $B(z,q_T)$ 
calculated at few fixed values of $q_T^2=0.5,\ 0.1$ and $0.2\GeV^2$.
Remarkably, the results of calculations demonstrate flattening and a drop of $B(z,q_T)$ towards $z=1$, similar to what is observed in data. Notice that these curves are not supposed to be directly compared with data, which cover an interval of $q_T$ dependent on $z$. The curves just show the uncertainty in the value of the slope, related to its ill definition.

\section{Determining the pion structure function from data}\label{F2pion}

Now we are in a position to try to answer the question, whether the process of leading neutron production in DIS can be consider as a tool to measure the pion structure function at small $x$. The answer depends on the sensitivity of the cross section to the number of quarks in the proton at a soft scale, and on the involved theoretical uncertainties.

If one trusted the way of calculations as is, then the number of quarks
$N_q^p$ would affect only the ratio (\ref{155nn}) and the cross section as a simple rescaling factor.
However, the procedure of interpolation of the absorption factor, Eq.~(\ref{499}), between the known regimes of short and long coherence length, Eq.~(\ref{496}), leaves some freedom in adjusting the shape of the $z$-dependence of the fractional cross section. The parameter $L$  is known only by the order of magnitude, it should be comparable with the nucleon size, i.e. $L\sim1\fm$. So far we fixed it at this value, however, after rescaling the cross section assuming different number of quarks $N_q^p$, we can readjust this parameter within a reasonable range in order to achieve a better agreement with data. 

\begin{figure}[!ht]
\centerline{
  \scalebox{0.35}{\includegraphics{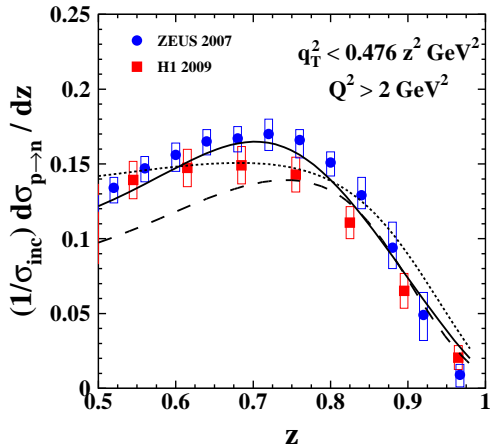}}
}
 \caption{(Color online) Data are the same as in Fig.~\ref{fig:qt-integrated}, upper panel.
 The dotted, solid and dashed curves are calculated with $N_q^p=3,\ 4,\ 5$, and with $L=2,\ 1$, and $0.5\fm$, respectively.}
 \label{fig:Np}
 \end{figure}

We considered two additional scenarios, which look to us extreme, $N_q^p=3$ and $N_q^p=5$, which correspond to no mesons, or in average to one meson in the light-cone wave function of the proton at a soft scale, respectively.
Correspondingly, we adjusted the length scale at $L=2\fm$ and $L=0.5\fm$ respectively.
The results a shown in Fig.~\ref{fig:Np} in comparison with data.
Apparently, the solid curve does the best job describing the data, and being optimistic one may conclude that the
version with $N_q^p=4$ is preferable.  However, being skeptical one may  say that the results of calculations are too model dependent to make a solid conclusion. Moreover, the relation between the ratio of pion to proton structure function and the mean number of quarks at a soft scale is based on the model of two scales in the proton \cite{spots2}, otherwise it may be broken.

\section{Summary}\label{summary}

To summarize, we highlight some of the results and observations.
\begin{itemize}

\item
Production of leading neutrons with fractional momentum $z\to1$ in DIS can serve as a way to measure the pion structure function at small $x$. However this method involves the several complications which are under investigation in this paper.

\item

We expect a reduction of the ratio $F_2^\pi(x)/F_2^p(x)$ compared with the usually assumed $2/3$. The deviation is due to the presence of a significant contribution of soft meson fluctuations in the proton wave function. While the contribution of the iso-vector mesons is constrained by the observed deviation from the Gottfried sum rule,  the role of the iso-scalar mesons in the proton is less known. The calculations performed in this paper are done with the fixed ratio $F_2^\pi(x)/F_2^p(x)=1/2$.

\item 
Even if the pion structure function is known, the fractional cross section of leading neutron production
cannot be accurately predicted because of initial/final state interactions generating absorptive corrections, whose magnitude is under debate.
Although the projectile particle in $\gamma^*p\to Xn$ is a highly virtual photon, we expect the effects of absorption
at large $z$, to be nearly as strong, as in $pp\to Xn$. This happens due to formation of a strongly interacting color0octet dipole $\{\bar qq\}_8^{\gamma^*}$-$\{\bar qq\}_8^{\pi}$.
The suppression factor $\tilde S_{4q}(b)$
depicted in Fig.~\ref{fig:S(b)} has a magnitude similar to $S_{5q}(b)$ in \cite{kpss,kpss-spin}.

\item The coherence time Eq.~(\ref{495}), which is the lifetime of the strongly interacting projectile color-octet dipoles,
is too short at medium values of $z$, and one should switch to the long-living $\bar qq$ fluctuations of the photon.
The corresponding absorption factor $S_{\gamma^*}(b)$ is evaluated to be rather close to one, as is depicted in Fig.~\ref{fig:S(b)}. The transition between the two regimes of absorption is illustrated in Fig.~\ref{fig:S(z)}.

\item
Other iso-vector meson exchanges, heavier than pion, are also important. The meson-nucleon couplings of natural 
parity $\rho$ and $a_2$ Reggeons, which are predominantly spin-flip, were fixed by phenomenological Regge fits to high-energy hadronic data.
The parameters of the unnatural parity $a_1$ Reggeon, which is non-spin-flip, are not well constrained by available data, and we fix them basing on the current algebra. Since the $a_1$ pole contribution was found to be very weak, we supplemented it by the $\rho$-$\pi$ cut, and treated them together as an effective $\tilde a_1$. Such an effective description was well tested in \cite{kpss-spin} with the
data on neutron azimuthal asymmetry.
The two sets of Reggeons have quite different intercepts and affect the neutron production cross section
in different regions of $z$. Fig.~\ref{fig:qt-integrated} shows that $\rho$ and $a_2$ are important at $z\to1$, while $\tilde a_1$ is large at smaller $z$.

\item
Eventually, we additionally tested two extreme assumptions about the number of quarks in the proton at a soft scale, $N_q^p=3$ and $5$. In each case we readjusted the parameter $L$ in Eq.~(\ref{499}) within a reasonable range, however, could not reach a good agreement with data, as is demonstrated in Fig.~\ref{fig:Np}.

\end{itemize}

Summarizing, our assumption Eq.~(\ref{final}) that the pion structure function at small $x$ is twice smaller than the proton one, is well supported by the parameter-free calculations of absorptive corrections and contribution of Reggeons,
providing a good description of ZEUS and H1 data for leading neutron production in DIS, as function of $z$, $Q^2$ and $q_T$. Nevertheless, we should admit that the test is not really precise because of many theoretical and experimental uncertainties involved into the calculations.

\begin{acknowledgments}

We are grateful to Armen Bunyatyan for 
very informative discussions of experimental results, and to Misha Ryskin for clarifications of the results of \cite{kkmr,kmr}. This work was supported in part
by Fondecyt (Chile) grants 1090236, 1090291 and 1100287, and by Conicyt-DFG grant No. 084-2009.
\end{acknowledgments}

\end{document}